\documentclass{article}[11pt]
\usepackage{fullpage}
\usepackage{hyperref}
\usepackage{url}
\usepackage{amsfonts}
\usepackage{xcolor}
\usepackage{graphicx}
\usepackage[capitalize]{cleveref}

\newcommand{\tool}[1]{\textcolor{black}{\textsl{#1}}}

\begin{document}

\begin{center}
{\Large{\bf Programming Frameworks for Differential Privacy\protect\footnote{To appear as a chapter in \cite{FiorettoVa24}.}}}\\*[3mm]
{Marco Gaboardi, gaboardi@bu.edu, Boston University\\
  Michael Hay, mhay@colgate.edu, Colgate University\\
Salil Vadhan, salil\_vadhan@harvard.edu, Harvard University} \\*[3mm]
\end{center}

Many programming frameworks have been introduced to support the development of differentially private software applications. In this chapter, we survey some of the conceptual ideas underlying these frameworks in a way that we hope will be helpful for both practitioners and researchers.  For practitioners, the survey can provide a starting point for understanding what features may be valuable when selecting a programming framework.  For researchers, it can help organize existing work in a unified way and provide context for understanding new features in future frameworks.  

We do not attempt to be comprehensive in our coverage of the landscape of software tools for differential privacy (which is constantly growing) or in the issues relevant to implementation.  In particular, we focus on programming frameworks for expressing and reasoning about differentially private statistical analyses, sometimes
referred to as ``queries.'' We refer readers interested in frameworks
for large-scale machine learning pipelines to Chapter~\ref{ch:7} and the references therein.   
Furthermore, there are a number of important
issues in programming with differential privacy that we did not address, such as randomness
generation, security, finite arithmetic, side channels, and scalability; discussions of these can be found in many of the papers we reference.  

\section{Introduction}
One of the reasons for 
the popularity of Differential Privacy~\cite{dwork2006calibrating,DworkMNS16} is 
that it is based on a
simple and elegant mathematical definition which enjoys several important
properties. 
One of these properties is {\em
  composability}, which is captured by a number of composition
theorems for differential privacy and its variants, as presented in Chapters~\ref{ch:1} and \ref{ch:3}.
In essence, these composition theorems tell us that the privacy
loss of a ``combined'' mechanism can be seen as a function of the
privacy losses of the different components. 
This enables reasoning in terms of 
a {\em privacy-loss budget}, where a global limit on privacy loss is enforced by tracking the accumulated privacy losses of differentially private queries made by data analysts. In addition, it also tells us that
different differentially private mechanisms can be put together to
form a ``combined'' mechanism incurring a graceful degradation of the
privacy loss.  This property naturally leads to the idea of
designing {\em programming frameworks} for differential privacy.
In this chapter, we use the term ``programming framework'' in a rather broad 
and inclusive way to refer to those tools that have been
designed to support the development of differentially private
applications and which provide some building blocks in the form of basic
mechanisms and methods to combine them. 

Composability is not the only property that makes the design of
programming frameworks possible. Another important property that
differential privacy enjoys is {\em resilience to postprocessing},
guaranteeing that manipulations of the result of a differentially
private mechanism cannot compromise the privacy of the data. (See Chapter~\ref{ch:1}.) In
addition, resilience to postprocessing also guarantees that once the
result of a mechanism is computed, we can forget about the method that
was used to compute it. This property provides a sort of {\em
  modularity} that is akin to the one usually encountered in programming
languages, where one can use library functions based on their
specification, without knowing the details of their implementation.
In particular, this property allows one to use differentially private
components inside larger software projects without worrying about how
the results computed by these components will be used.
Composability and resilience to postprocessing also permit one to
solve a data analysis problem by first decomposing it into subproblems 
and then combining the results.

There are several reasons why using programming frameworks for designing
differentially private applications is preferable to building
differential privacy applications from scratch:

\paragraph{Increasing reliability.}
Differential privacy is in its essence a probabilistic requirement on
programs and reasoning about probabilistic guarantees can be
challenging and error-prone, even for experts. (See, for example,
\cite{HaeberlenPN11,mironov2012significance,JinMcRuOh22,CasacubertaShVaWa22,HaneyDeHaShHa22}.) 
Errors in the design of
differential privacy analyses, or bugs in their implementations, can
compromise the privacy property of applications and undermine their
goals.  By providing a small set of programming primitives which can
be thoroughly vetted, programming frameworks can help improve
reliability and decrease errors in design and implementations of
differentially private data analyses.  In other words, programming
frameworks can reduce the verification of the privacy properties of
a program to the verification of the different components the frameworks provide.

\paragraph{Integration with familiar programming workflows.}
Programming frameworks for differential privacy provide programming-level
building blocks that can be used to implement differential privacy
applications. These building blocks are usually presented as
components of a library or of a domain-specific language. These
libraries and domain-specific languages can be integrated in standard
programming workflow. Hence, data analysts using these frameworks can
combine general-purpose programming and domain-specific programming.

\paragraph{Focusing on functionality and utility.} When designing a
differentially private data analysis for a specific statistical problem
one has to ensure that the analysis both respects the probabilistic requirement of differential privacy
and solves the intended statistical problem.  
Programming frameworks
often provide ways of automatically ensuring that the differential privacy guarantee is
met, without having to reason directly about the definition of differential privacy.
So, by using
these frameworks, a programmer or analyst can focus mainly on guaranteeing that the analysis
solves the statistical problem at hand, which is the domain expertise of most
end-user data scientists (who may not have expertise in differential privacy).

\paragraph{Supporting different computing environments.}
Some programming frameworks are designed in a way that decouples the
programming environment from the environment in which programs are
actually executed, often called the {\em computing environment}. This
separation allows a user to select a different computing environment based
on their own requirements, which can include efficiency, security, and/or 
integration with existing data infrastructure. 

\paragraph{Improving code reuse and helping build communities.}
Another benefit of programming
frameworks is that these frameworks can help the design of standardization
processes and open-source initiatives incentivizing code reuse and
better design practices. Programming frameworks can offer a common language
that can be used by different contributors as a {\em lingua franca}. This in
turn can help build communities of programmers and users around differential privacy.

\vspace{5mm}

There is by now a plethora of programming frameworks for differential
privacy.  A non-exhaustive list includes:
\tool{APEx}~\cite{ge2019apex},
\tool{Adaptive Fuzz}~\cite{winograd2017framework},
\tool{Airavat}~\cite{RoySKSW10},
\tool{Chorus}~\cite{JohnsonNHS20},
\tool{DFuzz}~\cite{gaboardi2013linear}
\tool{Diffprivlib}~\cite{holohan2019diffprivlib},
\tool{DPella}~\cite{lobo2020programming},
\tool{Duet}~\cite{NearDASGWSZSSS19},
\tool{Ektelo}~\cite{zhang2018ektelo},
\tool{Flex}~\cite{johnson2018towards},
\tool{Fuzz}~\cite{reed2010distance,HaeberlenPN11}
\tool{Google SQL}~\cite{WilsonZLDSG20},
\tool{Gupt}~\cite{MohanTSSC12},
\tool{OpenDP}~\cite{gaboardi2020programming},
\tool{PINQ}~\cite{mcsherry2009privacy},
\tool{PrivateSQL}~\cite{kotsogiannis2019privatesql}, and
\tool{PSI}~\cite{gaboardi2016psi}. 
%\sv{added Duet to above list and edited this text below}

Some of these are actively maintained open-source software tools, whereas others were only built as research prototypes.
Moreover, the landscape is rapidly changing, with new tools emerging at a rapid pace. Thus we do not attempt 
to provide a comprehensive description of any specific tools in this survey.  
Instead, we 
identify some of the essential characteristics that cut across
all of the tools, and only use the specific tools as exemplars of different design choices.

\paragraph*{Acknowledgements:} Marco Gaboardi's work was partially supported by NSF through award \# CNS-2040249.
Salil Vadhan's work was supported by a grant from the Sloan Foundation and a Simons Investigator Award. Marco Gaboardi and Salil Vadhan's work was also partially supported by Cooperative Agreement CB20ADR0160001 with the U.S. Census Bureau.  We thank
Gary Benedetto, Mark Fleischer, Philip Leclerc, and Rolando Rodr\'iguez from the Census Bureau for many helpful comments.
The views expressed in this paper are those of the authors and not those of the U.S. Census Bureau, or any other sponsor.

\section{Characteristics of DP Programming Frameworks}
\label{sec:programming-frameworks}

We organize our survey of differentially private programming frameworks around the following key
characteristics.

\paragraph{Privacy Calculus.} A specific aspect of programming frameworks 
for differential privacy is that they provide some 
form of a quantitative calculus that allows one to bound the privacy loss of analyses written
in the framework. 
These privacy calculi internalize different general techniques that have been
identified in the differential privacy research literature. For example,
several of the programming frameworks are based on privacy calculi designed around
the concept of global sensitivity~\cite{DworkMNS16}, and others around the sample-and-aggregate framework~\cite{NissimRS07}.
In most of the cases, these calculi provide a form of \emph{privacy specification} 
for different components that can be inspected to guarantee differential privacy 
for an entire analysis built from those components.

\paragraph{Composition and Interactivity.} 
In addition to providing a privacy calculus to determine the privacy loss of a single
analysis, programming frameworks for differential privacy often provide tools for tracking 
and controlling the cumulative 
privacy loss over multiple analyses via composition theorems for differential privacy. 
These tools range in their level of generality from supporting single, nonadaptive batches of
queries to interactive systems where a human analyst can adaptively choose the next query based
on the results of previous ones.

\paragraph{Expressivity.} One key design aspect of any programming framework is {\em expressivity}. Programming
frameworks for differential privacy are no exception. Differential privacy
is often presented as a property of programs working on some data, but
what exactly these programs are, and which statistical tasks they may be able to
accomplish, may vary. For example, some
programming frameworks are designed to support only programs representing
a restricted class of queries, while others attempt to
approximate arbitrary data analyses.

\subsection{Privacy Calculus}
\label{subsec:privacy-calculus}

Programming frameworks for differential privacy are usually based on some
form of a {\em privacy calculus}, which is a principled way to
guarantee that a data analysis is differentially private without
explicitly computing the probability distributions of the data
analysis on each pair of neighboring datasets.  Generally, the privacy
calculus specifies (a) how do we determine whether an individual
statistical computation is differentially private, and (b) how does
the privacy loss accumulate over multiple statistical computations.
The latter is known as {\em composition} and is treated in the next
section, along with interactivity.  In this section, we focus on the
former, how we determine that an individual computation is
differentially private.

Most of the differential privacy programming frameworks use some form of reasoning based on 
{\em sensitivity}, or equivalently, {\em stability}.\footnote{Historically, the terms sensitivity and stability have been used for related concepts in different settings in the differential privacy literature.  From the paper that defined differential privacy~\cite{DworkMNS16}, {\em sensitivity} has referred to functions mapping datasets to real numbers, vectors, or a general metric space, and measures how much the value of the function can change when one individual's data changes.  Starting from PINQ~\cite{mcsherry2009privacy}, {\em stability} has referred to functions mapping datasets to datasets, measuring how much output datasets can differ per change in output datasets. As discussed below, both are special cases of more general notions bounding output ``distances'' as a function of input ``distances'' for mappings between arbitrary sets that have some notion of elements being ``close'' to each other. We choose ``stability'' as the terminology for the more general notion, mainly because it represents a feature we seek: functions that are most compatible with privacy are ones that remain ``stable'' or ``{\em in-}sensitive'' to small changes in their input.} 

We describe these concepts and some of their different incarnations below.

\paragraph{Differential privacy and global stability.}
Recall from Chapter~\ref{ch:1} that a mechanism $M$ satisfies (pure) differentially privacy if for every two datasets $x,x'$ that are {\em adjacent} (i.e. differ on one individual's data),
the output distributions $M(x)$ and $M(x')$ are $\varepsilon$-close to each other (in the sense that for every set $T$ of outputs, the probabilities of $T$ under $M(x)$ and $M(x')$ are within a factor of $e^{\varepsilon}$ of each other).
More generally, by the group privacy property of pure differential privacy, if $x$ and $x'$ are at distance at most $d$ (i.e, differ on at most $d$ individuals' data), then the output distributions are at most $d\varepsilon$-close to each other.  Thus, differential privacy can be viewed as a {\em stability} or {\em Lipschitz} property of randomized algorithms $M$, whereby close inputs map to close output distributions.  We think of this as a {\em global} stability property because the stability constant $\varepsilon$ provides a uniform bound over all pairs of datasets $x$ and $x'$.

Thus, it is natural to use notions of global stability as the basis for the privacy calculus of a DP programming tool.  Below we describe some of the variants of global stability that arise in existing frameworks, in increasing levels of generality.

Before proceeding, we remark that, even for simple tabular datasets, there are many ways to formalize the notion of {\em adjacent} datasets, and it is important to be careful and consistent about this choice when implementing differential privacy.  In this chapter, except when explicitly stated otherwise, we model datasets as unordered multisets of records, and consider two datasets to be adjacent if we can obtain one from the other by adding or removing one individual's data. 
This formulation is often known as {\em unbounded DP} as it treats the number $n$ of records as being unbounded and potentially private information.  In contrast, {\em bounded DP} treats the number $n$ of records as fixed, public information, and two datasets as adjacent if one can be obtained from the other by {\em changing} one individual's data.

\paragraph{Stable dataset transformations.} The tool \tool{PINQ}~\cite{mcsherry2009privacy} supports {\em chaining} (pure) differentially private algorithms $M$ with transformations $T$ from datasets to datasets that are themselves stable.  Specifically, we say that $T$ is {\em $c$-stable} if for every two datasets $x$ and $x'$, the distance between the datasets $T(x)$ and $T(x')$ is at most a factor of $c$ larger than the distance between $x$ and $x'$. As a simple example, a ``map'' transformation 
\begin{equation}
    \label{eqn:map-transformation} T_f(\{x_1,\ldots,x_n\})=\{f(x_1),\ldots,f(x_n)\}
\end{equation} is 1-stable, since record $x_i$ of the input affects only one record $f(x_i)$ of the output.

Two key properties of stable transformations that make them useful for a privacy calculus are the following {\em Chaining Rules}:
\begin{enumerate}
    \item if $T$ is $c$-stable and $M$ is $\varepsilon$-DP, then $M\circ T$ is $c\varepsilon$-DP, and \label{itm:MTchaining}
    \item if $T_1$ is $c_1$-stable and $T_2$ is $c_2$-stable, then $T_1\circ T_2$ is $c_1c_2$-stable \label{itm:TTchaining}
\end{enumerate}
For example, using Chaining Rule~\ref{itm:MTchaining} with $T$ being a map function $T_f$ as in Equation~(\ref{eqn:map-transformation}) and $M$ being a differentially private noisy sum operator (which, say, clamps the values to $[0,1]$, adds them up, and adds Laplace noise), we can estimate sums of arbitrary bounded functions $f$ applied to records.

These two rules are the core of the privacy calculus of \tool{PINQ}~\cite{mcsherry2009privacy}.  In addition, \tool{PINQ} allows for reasoning about the stability of one-to-many transformations, which map a single dataset to several datasets (e.g. partitioning, which we will discuss in the next paragraph) and many-to-one transformations (e.g. database joins).

\newcommand{\cX}{\mathcal{X}}
\newcommand{\cY}{\mathcal{Y}}
\newcommand{\din}{d_{\mathrm{in}}}
\newcommand{\dout}{d_{\mathrm{out}}}
\newcommand{\eps}{\varepsilon}
\newcommand{\Lap}{\mathrm{Lap}}
\newcommand{\R}{\mathbb{R}}

\paragraph{Special cases of stable dataset transformations.}  Some frameworks are based on specific families of stable transformations.  For example, \tool{Airavat}~\cite{RoySKSW10} allows stable transformations of the mapping kind $T_f$ from Equation~(\ref{eqn:map-transformation}) as well as a $k$-stable generalization where each input record maps to $k$ output records:
$$T_{f_1,\ldots,f_k}(\{x_1,\ldots,x_n\})=\{f_j(x_i): j=1,\ldots,k, i=1,\ldots,n\}.$$
This allows for efficient implementation in a map-reduce framework. These map transformations can be composed with a small family of differentially private reducers, like the noisy sum.

A different generalization of map transformations is used, for example, in the programming tool
\tool{GUPT}~\cite{MohanTSSC12}, which allows $f$ to map a {\em set} of
records (i.e. a dataset) to a single record. Then we can obtain a
1-stable transformation by specifying  a (possibly random) partition $P=(P_1,\ldots,P_m)$
of the set of possible data records, and outputting
$$T_{P,f}(x)=\{f(x_{P_1}),\ldots,f(x_{P_m})\},$$
where $x_{P_i}$ consists of the records of $x$ in the partition piece $P_i$. This is a 1-stable
transformation because each input record $x_j$ affects at most one of
the output records.\footnote{
However, if we worked with bounded DP, then the transformation would only be 2-stable, because changing one input record could move that record from one piece of the partition to another, thereby affecting two of the output records.}
We can then apply an arbitrary differentially private
aggregator (such as a DP approximation of the average or median) to
the output of $T_{P,f}(x)$ to obtain a differentially private approximation of the estimator $f$, 
with the privacy justified by
Chaining Rule~\ref{itm:MTchaining}.  This realizes the
``sample-and-aggregate'' framework introduced in
\cite{NissimRS07}. The appeal of this framework is that $f$ can be an
arbitrary, {\em non-private} statistical analysis, and as long as $f$
gives similar results on the different subsets of the dataset
specified by $P$, then the aggregator should also give a similar
result (with additional noise, to achieve privacy).  The disadvantage
is that dataset sizes often need to be quite large for
sample-and-aggregate to be effective, since (a)
each set $P_i$ in the partition may need to be large in order for the
$f(x_{P_i})$'s to well-approximate $f(x)$, and (b) the number $m$
of sets on the partition may need to be large in order for the noise
introduced by the DP aggregator to be small.

\paragraph{Stability on general metric spaces.} Going beyond stable dataset-to-dataset transformations, \tool{Fuzz}~\cite{reed2010distance} can allow for reasoning about transformations between arbitrary metric spaces.  Specifically, a transformation $T : \cX\rightarrow \cY$ from metric space $(\cX,d_{\cX})$ to metric space $(\cY,d_{\cY})$ is {\em $c$-stable} (a.k.a. $c$-Lipschitz) if for all $x,x'\in \cX$, we have
$$d_{\cY}(T(x),T(x'))\leq c\cdot d_{\cX}(x,x').$$
For example, if $\cX$ is the space of all datasets, $d_{\cX}$ is the usual notion of distance on datasets discussed above, $\cY=\R$, and $d_{\cY}(y,y')=|y-y'|$, then a transformation $T$ is $c$-stable if and only if $T$ has {\em global sensitivity} at most $c$ --- the basis of the standard {\em Laplace mechanism} for differential privacy~\cite{DworkMNS16}.

A benefit of this generalization is that it allows for breaking differentially private algorithms into smaller components, which in turn provides more modularity, code reuse, and easier verifiability.  Let's consider the standard global sensitivity paradigm~\cite{DworkMNS16} 
for designing differentially private algorithms.  Let $T$ be a function from datasets to real numbers that has global sensitivity at most $c$, i.e. is $c$-stable as above. Then, as explained in 
Chapter~\ref{ch:1}, the mechanism 
$$M(x) = T(x)+\Lap(c/\eps),$$ is an $\eps$-differentially private
algorithm.  With reasoning about stability between arbitrary metric
spaces, we can decompose the construction and analysis of $M$ into two simpler, modular components.
Specifically, note that $M=N\circ T$, where
$T$ is the function of global sensitivity at most $c$, and
$N(z)=z+\Lap(c/\eps)$ adds noise to single real numbers.  The noise-addition mechanism 
$N$ has the
property that two inputs $z$ and $z'$ that are close as real numbers
(rather than as datasets) map to close output distributions.  Specifically the distance between
the distributions of $N(z)$ and $N(z')$ is at most a factor of $(\eps/c)$ times $|z-z'|$.  Thus,
by an appropriate generalization of Chaining Rule~\ref{itm:MTchaining}, the composition $N\circ T$
is $c\cdot (\eps/c)$-DP.

More
generally, we call a (randomized) function like $N$ where close inputs (with respect to
any input metric) map to close output distributions (with respect to
any metric on probability distributions) a {\em
  measurement}. 
In contrast, if we only allowed for reasoning about stable
transformations and DP algorithms on datasets as above, then $M=T\circ
N$ would have to be treated as an atomic function; now we can separate
the components, and reuse the same noise-addition measurement $N$ for different
stable transformations $T$ (and conversely).

Another benefit of this generalization is extending the privacy calculus to other input types, such as a database consisting of multiple tables.  
What distance metric is appropriate for a database is a nuanced question that depends on what data is being represented.  Consider, for instance, a graph database consisting of a $\textrm{Node}$ table and an $\textrm{Edge}$ table.  The right choice of metric would depend on whether one wants to protect a single edge, a node and all of its incident edges, or some other property of the data.
Several programming frameworks support database inputs but vary in the distance metrics they support. For example,  \tool{Flex}~\cite{johnson2018towards} measures distance in terms of number of rows (across all tables).  On the graph database example, whether Flex offers node or edge-level protection would depend on the query being asked.  \tool{PrivateSQL}~\cite{kotsogiannis2019privatesql} uses distance metrics that incorporate integrity constraints that hold on the data.  An example of such a constraint is if $u$ is removed from the Node table, then all incident edges must be removed from the Edge table as well.  This allows \tool{PrivateSQL}, for instance, to capture node-level protection.
\tool{GoogleSQL}~\cite{WilsonZLDSG20} is designed for databases where each row is ``owned'' by a user, each user may own several rows, and distances between two databases is measured in users rather than individual rows. 
This captures the notion of user-level protection that is desirable in applications where one user might ``own'' a large number of rows (e.g., a database of credit card transactions). A limitation of this approach, however, is that it cannot be easily applied to graph data, because it cannot support scenarios where a data item can have multiple owners, which is the case for edges in this example.\footnote{For more information on differential privacy for databases, we recommend Near and He~\cite{near2021differential}.}

\paragraph{Beyond metric spaces.} The standard mathematical definition of a {\em metric} $d(x,x')$ is a real-valued function that satisfies requirements of nonnegativity, symmetry, and the triangle inequality. However, not all ``closeness'' notions that appear in the design of differentially private algorithms are convenient to express in terms of such metrics.  For example, {\em approximate differential privacy} measures the closeness of distributions via two real parameters $\varepsilon$ and $\delta$, whereas metrics are required to express distance via a single real number.  Thus, the \tool{OpenDP} Programming Framework~\cite{gaboardi2020programming} allows for expressing more general stability notions whereby ``$\din$-close'' inputs map to ``$\dout$-close'' outputs (or output distributions), for arbitrary, user-defined closeness relations that can be expressed in terms of parameters $\din$ and $\dout$ of arbitrary type.  In particular, \tool{OpenDP} can support a variety of privacy measures, such as approximate DP~\cite{DworkKeMcMiNa06}, concentrated DP~\cite{DworkRo16,BunSt16}, and 
$f$-DP~\cite{DongRoSu22}.
Also \tool{Duet}~\cite{NearDASGWSZSSS19} supports several privacy measures but it only supports stability notions over metric spaces. \tool{Duet} combines two domain-specific languages: one is similar to \tool{Fuzz} and supports only reasoning about transformations between arbitrary metric spaces, while the other supports instead reasoning about arbitrary measurements. 

\newcommand{\LS}{\mathrm{LS}}

\paragraph{Local stability.}  A difficulty with reasoning about global stability is that it can sometimes be overly conservative, in that the pair $x,x'$ of datasets that maximize $d(T(x),T(x'))/d(x,x')$ may be artificial ones that are unlikely to arise in practice.  Thus, it is attractive to try to reason using {\em local stability}, where we take $x$ to be our actual, given dataset, and consider the quantity
$$\LS_T(x) = \max_{x' : d(x,x')=1} d(T(x),T(x')).$$  Note that this is a data-dependent quantity, in contrast to global stability that does not depend on the given $x$.

Unfortunately, directly trying to substitute local stability for global stability in DP algorithms generally does not preserve privacy; the problem is that the stability itself (which ultimately controls the amount of noise introduced) can reveal sensitive information.  Thus, there is a large body of work on how to approximate this idea (using variants of local stability) while obtaining algorithms that are genuinely differentially private.  
\tool{Flex}~\cite{johnson2018towards} is a tool whose design draws on this body of work to tackle the problem of privately answering database queries that involve joins.  
Whereas a join query can have large, possibly unbounded, \emph{global} stability, its \emph{local} stability can be much smaller.  For instance, local sensitivity of a join between a $\textrm{Node}$ and $\textrm{Edge}$ tables would be a function of the maximum node degree as opposed to the number of nodes.
\tool{Flex} leverages this property in its approach, called \emph{elastic sensitivity}, which is a calculus for deriving an upper bound on the local stability of a complex database query expression.  \tool{Flex} combines its elastic sensitivity with techniques from prior work on appropriately calibrating noise to bounds on local stability.
\subsection{Composition and Interactivity} \label{subsec:composition}

In differential privacy, \emph{composition} 
refers to reasoning about the privacy properties of an algorithm that is composed of one or more differentially private algorithms.  
The basic composition theorem for (pure) differential privacy says that if mechanisms $M_1$ and $M_2$ are $\epsilon_1$-DP and  $\epsilon_2$-DP respectively, then the mechanism $M(x)=(M_1(x),M_2(x))$ that applies both mechanisms to the same dataset and
outputs both results is $(\epsilon_1 + \epsilon_2)$-DP.
Composition has two key applications in the context of programming frameworks.  First, it allows for tracking the \emph{cumulative} privacy loss that results from executing multiple statistical computations.  Some programming frameworks help users manage the cumulative privacy loss by allowing them to impose a privacy loss ``budget'' and ensuring that the cumulative loss does not exceed the budget.  
Second, it is a useful tool in algorithm design as it facilitates the creation of new differentially private algorithms from existing components.  Some programming frameworks facilitate the creation of new mechanisms in this way.  Both of these applications of composition are enhanced by frameworks that support \emph{interactivity}, where human analysts or programs can adaptively choose which mechanism $M_2$ to apply after receiving the result of a previous mechanism $M_1$.

While the composition rules for pure DP are fairly straightforward (epsilons add up), the composition rules for other variants of DP are less so, and this has been and continues to be an active area of research~\cite{kairouz2015composition,murtagh2016complexity,RogersRoUlVa16,abadi2016deep,DworkRo16,BunSt16,mironov2017renyi,
DongRoSu22,vadhan2021concurrent,vadhan2022concurrent,lyu2022composition,whitehouse2022fully}.  (See Chapter~\ref{ch:3} for more on composition.)
We organize the rest of this section around the kinds of composition that are supported by programming frameworks for differential privacy.

\paragraph{No composition.} Some programming frameworks, such as \tool{Flex}~\cite{johnson2018towards} and \tool{Diffprivlib}~\cite{holohan2019diffprivlib}, 
do not support composition or interactivity. That is, they enable one to control or calculate the privacy loss of an individual differentially private mechanism or query, but tracking the cumulative privacy loss is left to the user or an application
built on top of the framework. 

\paragraph{Non-adaptive composition.} 
Several programming frameworks allow the user to construct a program that evaluates a \emph{batch} of statistical computations $M_1, \dots, M_{\ell}$.  
The computations are \emph{nonadaptive}, meaning that the sequence of computations is fixed: each $M_i$ has fixed privacy loss parameters and its only input is the private dataset.   An archetypal example is the released prototype of \tool{PSI}~\cite{gaboardi2016psi} and its successor \tool{DP Creator}~\cite{DPCreator22}, which allows a data curator to choose a set of summary statistics (e.g., means, quantiles, histograms) to be released under a fixed privacy loss budget.  Other frameworks that fall into this category include \tool{Airavat}~\cite{RoySKSW10}, \tool{PrivateSQL}~\cite{kotsogiannis2019privatesql}.

Such frameworks are responsible for calculating the cumulative privacy loss of the program and typically ensuring it does not exceed a user-specified budget.  
Some employ the basic, pure DP sequential composition theorem alluded to in the introduction to this section, but \tool{PSI} uses (an approximation of) the optimal composition theorem for approximate DP~\cite{murtagh2016complexity}.
However, these frameworks do not explicitly track the cumulative privacy loss across batches or support interaction with the user.\footnote{The paper describing the design of \tool{PSI}~\cite{gaboardi2016psi} proposes a way to support interactivity with multiple batches, but the prototype implementation does not offer that functionality.}

\paragraph{Adaptive composition.} Adaptive composition refers to the idea that subsequent mechanisms may depend on the results of evaluating earlier mechanisms ($M_2$ can depend on $M_1(x)$).  In the most basic form of adaptive composition, one again has a fixed sequence of $\ell$ statistical computations, $M_1, \dots, M_\ell$ where the privacy loss of each $M_i$ is fixed, but each subsequent computation can depend on the results of previous computations---i.e., mechanism $M_i$ is provided the results of earlier computations as auxiliary input.  Formally, we compute the $i$'th result as $y_i = M_i(x ; (y_1,y_2,\ldots,y_{i-1}))$, where $x$ is the private dataset. 
With this kind of adaptivity, one can support iterative algorithms such as $k$-means clustering, an algorithm for grouping data into $k$ clusters.  Such an algorithm is inherently adaptive because it proceeds in rounds where the assignment to clusters in each round depends on a statistical summary produced in the previous round.

An example of a programming tool that supports this kind of composition is \tool{Fuzz}.  \tool{Fuzz} uses a type system to statically analyze (at ``compile time'') the composition of a program.  While \tool{Fuzz} does support adaptive composition, it is limited to programs where the stability (or sensitivity, see \cref{subsec:privacy-calculus}) is a constant and therefore 
precludes supporting, say, a version of $k$-means where the number $k$ of clusters or the number $\ell$ of iterations are user-specified inputs to the program.  \tool{DFuzz}~\cite{gaboardi2013linear} overcomes this limitation by extending the static type system approach to allow types to depend on input variables. Other more complex type system have been studied, in order to go even farther in supporting the verification of differential privacy data analyses. We will discuss some of them briefly in \cref{subsec:verification}. 

\paragraph{Fully adaptive composition and interactivity.} 
With the previously described programming frameworks, the entire statistical computation, even if adaptive, must be specified up front.  While such frameworks can be useful for building complex algorithms, in many practical applications, however, it may be undesirable to bundle all computations into a single ``one shot'' algorithm. Instead, analysts may prefer to interactively query the data, receive (differentially private) responses, and adaptively choose the next query.  In such a setting, the composition can be said to be fully adaptive because the number of rounds, the choice of mechanism, and its privacy parameters are all adaptively chosen by the analyst.  Composition is more subtle in this setting~\cite{RogersRoUlVa16}.

The interactive setting can be modeled as an interactive protocol between two parties, an arbitrary adversarial strategy $\mathcal{A}$ (the analyst) and an interactive measurement $\mathcal{M}(x)$.  The interaction proceeds in rounds where in round $i$, $\mathcal{A}$ selects a query $q_i$ adaptively based on all previous answers $(a_1, \dots, a_{i-1})$ and any randomness of $\mathcal{A}$.  One can think of a query as a request to perform a differentially private computation with given privacy loss parameters.
The query is sent to the interactive mechanism $\mathcal{M}$ which evaluates the query and returns answer $a_i$.  
Differential privacy is defined with respect the adversary's \emph{view}, denoted $\mathrm{View}(\mathcal{A} \leftrightarrow \mathcal{M}(x))$, of this interaction~\cite{vadhan2021concurrent}.

We highlight two programming frameworks in this setting.  Both support a form of interaction known as a \emph{privacy filter}~\cite{RogersRoUlVa16}, an interactive mechanism that takes a global privacy loss budget and admits adaptively selected computations provided that evaluating the computation would not cause the cumulative privacy loss to exceed a pre-specified budget.  A variant is the concept of a \emph{privacy odometer}~\cite{RogersRoUlVa16}, which tracks cumulative privacy loss but does not enforce a pre-determined budget. 

In the programming tool \tool{PINQ}, the interactive mechanism is represented by a \emph{PINQueryable}, a stateful object that encapsulates a private data source (a collection of records) and allows it to be queried within a given privacy-loss budget. 
Each \emph{PINQueryable} exposes a set of pre-defined methods, which represent the available queries.  An example method is $\mathrm{NoisyAverage}$, 
which takes an $\epsilon$ and a function $f$ and applies $f$ to each tuple in the source, clamps the result to $[-1,1]$, and returns a noisy average of the clamped results.
The analyst $\mathcal{A}$ can be expressed as arbitrary code, written in the language C\#; in particular, $\mathcal{A}$ can make fully adaptive queries to the \emph{PINQueryable}, and the \emph{PINQueryable} tracks the cumulative privacy loss and enforces the budget.

The programming tool \tool{Adaptive Fuzz}~\cite{winograd2017framework} is built on top of \tool{Fuzz}: it adds an outer ``adaptive'' layer that wraps around the static typechecker and runtime of \tool{Fuzz}.  
Unlike \tool{PINQ}, the available queries are not limited to a pre-defined set of methods, but rather a query can be any computation that the \tool{Fuzz} typechecker can verify to be differentially private.
An example of an application that \tool{Adaptive Fuzz} can support is fitting a model using differentially private gradient descent with a data-dependent stopping criteria, such as the model error falling below a threshold.  If the stopping criteria is reached before the budget is exhausted, any remaining budget can be applied to additional (adaptively chosen) computations.

\paragraph{Hierarchical (and concurrent) interactivity.} While there are several frameworks that support interactivity, most operate like \tool{Adaptive Fuzz} with a single ``level'' of interactivity: there is exactly one interactive mechanism that governs the adaptivity and any query to that mechanism must correspond to a  \emph{non-interactive} differentially private program.  \emph{Hierarchical interactivity} refers to the idea that an interactive mechanism could recursively spawn new interactive mechanisms.

Hierarchical interaction has many potential applications.   It could be useful as a way of structuring programs that naturally have multiple layers of interactivity -- for instance, the first layer of interactivity might be a privacy filter and the second layer of interactivity might include the execution of interactive primitives like Sparse Vector or Private Multiplicative Weights~\cite{dwork2014algorithmic}. 
Another application could be to allow a data curator to allow \emph{multiple} analysts to operate concurrently on the data, each with their own privacy filter or odometer, such that the cumulative privacy loss is tracked appropriately.  
One could also imagine novel algorithms that involve asking interleaving queries to two or more instantiations of interactive mechanisms.

\begin{figure}[ht]
\centering
\includegraphics[width=.9\textwidth]{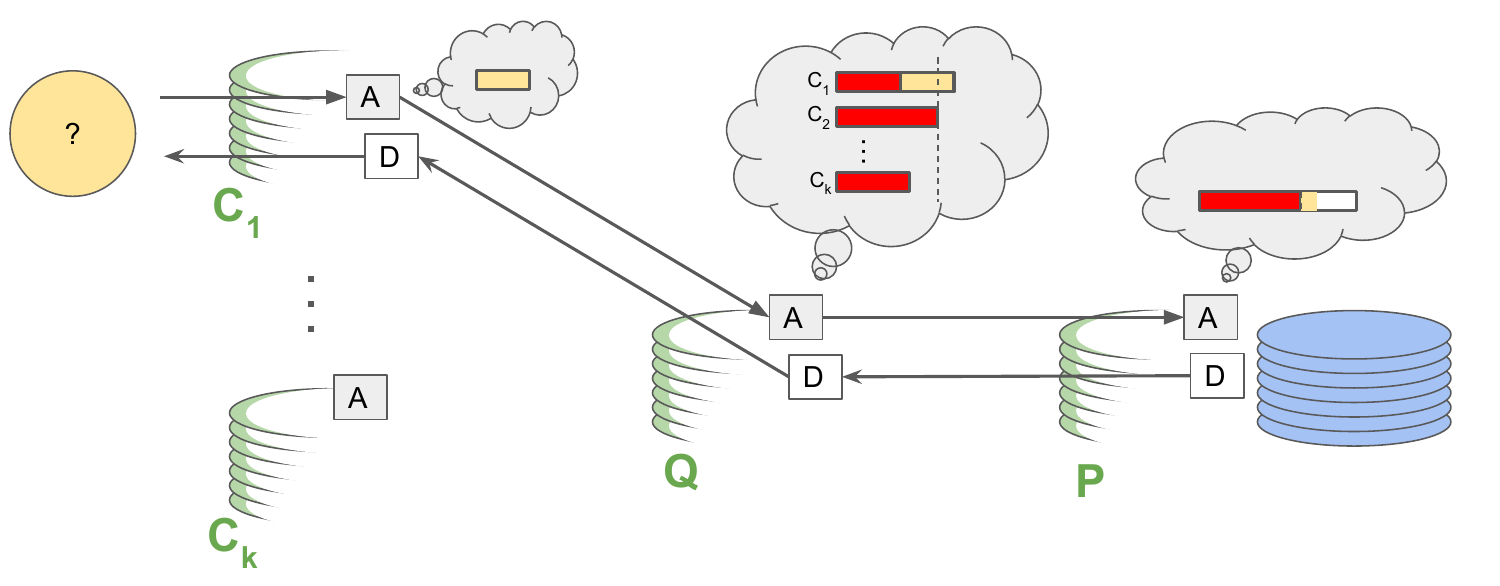}
\caption{\label{fig:PINQ} Illustration of hierarchical interactivity in \tool{PINQ}.  When a query (yellow circle) is submitted to \emph{PINQueryable} $C_1$, its agent (A) forwards the budget request (yellow bar) to the agent of its parent, $Q$.  In this case, because the children represent disjoint partitions of $Q$'s data, the agent of $Q$ keeps track of the cumulative privacy loss incurred at each child (red bars). When it receives the request from the child, it calculates how much this query would increase the maximum privacy loss and sends a request to the agent of its parent, $P$, who checks the residual cost against a budget.  Since the request is under budget, response data (D) flows back to the \emph{PINQueryable} where the query initiated.}
\end{figure}

\tool{PINQ} supports (a limited form of) hierarchical interactivity.  Some of the methods available on a  \emph{PINQueryable} are capable of spawning new \emph{PINQueryable} objects.  For instance, the $\mathrm{Partition}$ method splits the dataset into multiple, disjoint datasets and returns a new \emph{PINQueryable} for each one.  Using such methods it is possible to start with an initial private data source and spawn a tree of related queryables.\footnote{In fact, rather than a tree, the result structure can be a directed acyclic graph because \tool{PINQ} also supports operators like $\mathrm{Join}$ that combines two queryables into one.}  Such a tree is illustrated in~\cref{fig:PINQ}.

\cref{fig:PINQ} also illustrates \tool{PINQ}'s method for reasoning about the the composition of these multiple, related queryables.  Each queryable internally has an agent which is responsible for managing requests to consume privacy budget through one of the measurement methods (like NoisyAverage).   The agent associated with each new queryable has a reference to its parent queryable. The child is responsible for handling its request and translating it into an appropriate request to its parent.  For instance suppose $Q$ is a queryable with parent $P$ and  children $C_1, \dots, C_k$ that are the result of calling $\mathrm{Partition}$ on $Q$.  Because partition splits the data into disjoint subsets, interactions with the children satisfy \emph{parallel composition}~\cite{mcsherry2009privacy} and the cumulative privacy loss is the maximum of the loss incurred by any of the children.\footnote{This is similar to the stability analysis underlying the sample-and-aggregate framework and GUPT discussed above in \cref{subsec:privacy-calculus}.} 
Therefore, when the agent of $Q$ receives a request from its children, the amount of the request to its parent is only the \emph{change} in the maximum (which might be zero).

Somewhat surprisingly, the more general concurrent composition of multiple mechanisms has only recently been formally studied~\cite{vadhan2021concurrent} and 
researchers have only recently presented proofs that other variants of DP such as $f$-DP~\cite{DongRoSu22} and Renyi-DP~\cite{mironov2017renyi,BunSt16} satisfy concurrent composition theorems~\cite{vadhan2022concurrent,lyu2022composition,HaneyShTiVaVyXuZh23}.

\paragraph{Extensibility.} With both \tool{PINQ} and \tool{AdaptiveFuzz}, the composition is ``baked in'' to the programming framework itself.  For example, the implementation of PINQ supports a specific definition of privacy (pure DP); its components exchange information using the parameters of that definition ($\epsilon$) and the agents apply the composition rules of PureDP.   
 Perhaps more fundamentally, it leans heavily on the fact that with group privacy under pure DP, the privacy loss grows linearly with the size of the group, which allows it to view an $\eps$-DP measurement on a $c$-stable transformation of the dataset as equivalent to a $c\eps$-DP measurement on the source.  (See \cref{subsec:privacy-calculus}.) Adapting PINQ to support a variant of DP like zCDP that has a non-linear relationship between group size and privacy loss seems non-trivial, especially if one wants to provide tight privacy accounting.

An explicit design goal of \tool{OpenDP} is extensibility and its approach to handling composition makes it easy, in principle, to expand the library with new forms of composition.  In the OpenDP formulation of a {\em measurement} (as discussed in \cref{subsec:privacy-calculus}), the privacy properties are explicitly encoded in a {\em privacy map} that transforms distances between input data(sets) to distances between output probability distributions, and the latter can be measured in any desired way.

To extend OpenDP with a new form of composition, one must simply define a new measurement operator that captures that composition operation.  For example, basic non-adaptive composition is implemented as a
measurement function that takes as input a list of measurement functions and applies them sequentially to the data.  It can use the privacy maps of the measurements being combined plus the sequential composition theorem to determine its own privacy map.

\subsection{Expressivity} 
Expressivity refers to the richness and variety of analyses that can be expressed in a given programming framework. 
These are naturally impacted by a framework's privacy calculus and support for composition.  In particular, given that most analyses are expressed as a chaining of measurements and stable transformations, e.g. $M\circ T_k\circ T_{k-1}\circ \cdots T_1$, or compositions of such chainings, the expressivity of a programming framework is greatly impacted by which transformations, measurements, and composition procedures are supported.
Thus, another relevant feature is {\em extensibility}, the support for adding new, user-defined building-block components such as
transformations, measurements, and composition procedures, as just discussed.
Following the same approach we used in previous sections we will not aim to have a general ranking between the frameworks in terms of expressivity, rather we will highlight some points with respect to expressivity in the design space of programming frameworks for differential privacy. 

\paragraph{Allowing general-purpose programming together with DP.} Several programming frameworks for differential privacy have been designed as libraries or domain-specific languages embedded inside general-purpose programming languages.  This allows one to use the richness of the host language to enhance the expressivity of the programming framework. One of the earliest such examples is \tool{PINQ}~\cite{mcsherry2009privacy}, which is designed as a domain-specific language integrated in the general-purpose language C\#. Specifically, \tool{PINQ} has been implemented as a C\# component built on top of the data-processing component LINQ. \tool{PINQ} provides several basic transformations, mostly inspired by SQL such as filters, joins, groupby, map, etc and several basic measurements representing aggregation operations, such as count, sum, average, etc. In fact, most of the \tool{PINQ} transformations and measurements consists of a thin layer of code encapsulating the corresponding LINQ transformations. This encapsulation mechanism is needed to guarantee differential privacy through a privacy calculus based on global sensitivity of the sort we discussed in \cref{subsec:privacy-calculus}.

Thanks to the integration with a general purpose programming language, \tool{PINQ} supports general purpose programming that can be used to describe more complex data types, and to implement new transformation and mechanisms.
Indeed, 
\tool{PINQ}'s set of data types, transformations and measurements can be easily extended, in principle, by providing new implementations that match the same interface that primitive components have to match, and that respect the principles of the privacy calculus that \tool{PINQ} implements. In fact this is a recipe for extensibility that several other programming frameworks for differential privacy have followed: new transformation and measurement primitives can be easily added to a programming framework by taking operations that are implemented in the underlying language and encapsulating them in order to guarantee their stability and sensitivity properties for their sound use in the corresponding privacy calculus.  

As we discussed in \cref{subsec:composition},  private data sources in \tool{PINQ} are encapsulated in a \emph{PINQueryable} object and can be accessed  through chains of transformations followed by an aggregation. This chaining is one of the main principles for the interactions of \emph{PINQueryable} objects with transformations and measurements. The encapsulation of data using queryables allows, in principle, extensions of \tool{PINQ} that implement different forms of composition with different degrees of adaptivity and hierarchical and concurrent interactivity.  Both chaining and composition are implemented in \tool{PINQ} as C\# routines that offer at the same time an interface for users and a component of the privacy calculus. Similarly to transformations and measurements, one could extend this family of chaining and composition components by providing new implementations in the underlying general-purpose language.
However, while possible in principle, as discussed in the \tool{PINQ} design document~\cite{mcsherry2009privacy}, these extensions are less straightforward than extensions of transformations and measurements, since they require redefining \emph{PINQueryable} and its methods. More recent programming frameworks such as \tool{OpenDP}, take a more explicit view on chaining and composition components: they are first class citizens,
called {\em combinators}, with an explicit interface that can be used to interact with them, and explicit requirements for their privacy calculus properties. Thus, to add a new form of composition to \tool{OpenDP} simply requires adding a new combinator capturing it. 

In some situations, programming frameworks for differential privacy also allow user to write and run arbitrary programs. Often these programs are required to be pure or ``purified,'' i.e. without side effects that can potentially compromise privacy. Programming frameworks that work in synergy with a general-purpose programming language can use the resources of the host language in these situations. For example, as discussed in the \tool{PINQ} design document~\cite{mcsherry2009privacy}, one can write C\# programs using a set of methods which are side-effect free and which can be used in mapping operations in a data processing. The distinction between methods that can have potential side effects and those that are side-effect-free is built into \tool{PINQ}.
Other approaches, such as the one implemented in DPella~\cite{lobo2020programming},  make this kind of distinction using types. 

There is one additional benefit of using a general-purpose language that is worth discussing. General-purpose programming languages usually provide methods to access a variety of different kinds of data sources, e.g. relational databases, tabular data in various forms, streaming sources, etc. 
This allow programming frameworks for differential privacy to easily also support different kinds of data sources, and to process data independently from the way it is stored.

\paragraph{Support for a specific class of queries.} Most of the programming frameworks, instead of aiming to support the implementation of arbitrary differentially private data analyses,  focus on a specific class of functions that are useful to implement a large class of analyses. For example, \tool{PINQ}~\cite{mcsherry2009privacy}, \tool{PrivateSQL}~\cite{kotsogiannis2019privatesql}, \tool{Flex}~\cite{johnson2018towards}, and others all aim at supporting SQL-like queries. 

A different example is \tool{Ektelo}~\cite{zhang2018ektelo}. Essentially, \tool{Ektelo} is very similar to \tool{PINQ} in many respects: it also provides transformations and measurement similar to the ones provided by \tool{PINQ}, and a privacy calculus for them. However, instead of focusing on the components that are needed for SQL-like queries, it includes a large variety of basic transformations and measurements that allow one to construct numerous important differentially private algorithms for estimating workloads of {\em linear queries}. Linear queries are queries of a simplified form: they compute the average, or the sum, of the results obtained by mapping  a given function (characterizing the specific linear query) on each record of a dataset. 
Despite their simplicity, when combined together in a workload, i.e. multiple queries together, linear queries allow one to solve many important statistical problems, such as histograms, CDFs, and contingency tables.

\tool{Ektelo}'s basic components include transformations and measurements, but also operations of three other families: 1) Operations that allow one to partition the data and to work on the individual partitions. For example, \tool{Ektelo} provides a data-dependent operation {\tt AHPpartition} which, given the data as an histogram, selects a partition of the data where the counts within a partition group are close. 2) Operations that allow one to manage the workload of queries by selecting queries that have different properties. For example,  \tool{Ektelo} supports an operation {\tt Worst-approx} which returns a query, from a workload of queries, which has the property that
on a synthetic dataset (or synthetic data distribution) deviates most from the true dataset. 
3) Operations that perform some sort of inference as post-processing of the private answers. For example, \tool{Ektelo} provides an operation {\tt MW} which implements the Multiplicative Weight inference algorithm. All these operations can be combined to design more complex data analyses. For example, one can use basic transformations and measurements and operations of types 2) and 3) to implement the private data release algorithm MWEM~\cite{hardt2012simple}.

Despite the fact that its design focuses on the restricted class of linear queries, \tool{Ektelo} is able to express in a concise way numerous differentially private algorithm that have been proposed for a variety of tasks. Moreover, thanks to its additional operations, \tool{Ektelo} is also able to go beyond tasks that are usually considered as proper linear queries. For example, one can use \tool{Ektelo} to  privately select the structure of a Bayesian network. 

Another programming framework using an approach similar to the one of \tool{Ektelo} is \tool{Chorus}~\cite{JohnsonNHS20}. Instead of focusing on linear queries, \tool{Chorus} focuses on SQL queries and it provides several components to rewrite, analyze and post-process queries. These components allow \tool{Chorus} to express in a concise way numerous complex differentially private algorithm such as MWEM, Sparse Vector, etc.

\paragraph{Support for a specific class of DP algorithms.} As we discussed in \cref{subsec:privacy-calculus}, there are several privacy calculi that one could use to guarantee differential privacy. Programming frameworks usually support one such privacy calculus, and the privacy calculus that a programming framework uses also affects its expressivity. 

Programming frameworks that have been designed around the notion of global stability, such as \tool{PINQ} and many others, allow one to implement a broad range of general differentially private data analyses. In contrast, programming frameworks that support other models, such as \tool{GUPT}~\cite{MohanTSSC12} for the sample and aggregate framework, or \tool{Flex}~\cite{johnson2018towards} for the local stability framework, are more limited in the class of data analyses that they can support. 

The limitations of \tool{GUPT}, and the sample and aggregate framework more generally, come from the fact that this framework only captures differentially private data analyses that can be implemented as a transformation that can be run independently on the pieces of a partition of the dataset, followed by one of a few pre-specified differentially private aggregators combining the obtained results. 
Most differentially private algorithms cannot be decomposed in this way.

The limitations of \tool{Flex} and other local stability frameworks is that these frameworks are usually based on some approximation of the local stability that cannot in general be computed efficiently for every problem. This means that the programming framework using these approximation needs to be specialized to specific tasks. For example, \tool{Flex}'s  elastic sensitivity calculus is based on a careful case analysis of operators that appear in SQL expressions. 
It is not obvious how to extend this definition to arbitrary data analyses, and it is unclear whether such an extension would be tractable. 

While these frameworks have limitations in terms of expressivity, they also have some benefit. For example, as we already mentioned in \cref{subsec:privacy-calculus}, the sample-and-aggregate framework
allows one to take arbitrary non-private estimators that converge as the sample size grows and automatically construct differentially private analogues of them, albeit on larger datasets. Similarly, methods based on local stability can provide improved accuracy in certain situations.

\paragraph{Support for reasoning about accuracy.}
Most of the programming frameworks we discussed so far focus on providing support to users in order to write their differentially private data analyses and guarantee they are indeed private. However, another important aspect that users need to think about when designing differentially private analyses is \emph{accuracy}. While there isn't one notion of accuracy which works in every situation, a notion of accuracy which is often considered in the differential privacy literature is based on probability bounds. 
As an example, $(\alpha, \beta)$-accuracy expresses the following probability bound: 
if $\tilde{y}$ is the result of a differentially private analysis and $y$ is the result one would achieve if one did the analysis without privacy constraints, then $(\alpha, \beta)$-accuracy says that with probability at least $1-\beta$, 
$\| y - \tilde{y} \| \leq \alpha$ where $\| \cdot \|$ is a suitable norm.
For simple differentially private mechanisms, the $\alpha$ and $\beta$ follow from the properties of the noise distributions they use.

Several programming frameworks have been conceived to also help users reason about the accuracy of their differentially private data analyses. One such example is \tool{GUPT}~\cite{MohanTSSC12}, which provides its users with the possibility to select either the privacy budget or the $(\alpha,\beta)$-accuracy they want for each data analysis. If users specify the accuracy level they are interested in, \tool{GUPT} computes the corresponding privacy budget that is needed, and vice versa: if users specify the privacy budget they are interested in, \tool{GUPT} computes the corresponding $(\alpha,\beta)$-accuracy.
A similar approach is also used by \tool{PSI}~\cite{gaboardi2016psi}, where users are provided with a budgeter interface that they can use to specify interactively either the privacy budget or the accuracy for the different queries.
The usability of this budgeter interface has also been evaluated with users~\cite{MurtaghTKV18,SarathySoHaScVa22}.
The approach used by \tool{GUPT} and \tool{PSI} is based on the accuracy bounds of the primitive noise distributions they use. \tool{APEx}~\cite{ge2019apex} takes a further step and uses several specific analyses and linear programming to provide accuracy estimates for more complex algorithms, including some algorithms that have accuracy which depends on the data. 

A different approach is proposed by \tool{DPella}~\cite{lobo2020programming}, a programming framework which combines reasoning about privacy in the style 
of \tool{PINQ},\footnote{Differently from what happen in \tool{PINQ}, the privacy reasoning in DPella happens at compile time.} with reasoning about $(\alpha,\beta)$-accuracy. A user of \tool{DPella} can write differentially private programs and obtain information about their privacy or accuracy. An interesting aspect of the \tool{DPella} approach is that it allows to compose the accuracy of different mechanisms using the Union bound and Chernoff bounds in order to provide accuracy properties for larger programs.

\section{Tools for Verification and Testing}
As we discussed above, an important motivation for the design of
domain specific programming frameworks for differential privacy comes
from the difficulties that randomness, finite precision arithmetic, budget accounting, etc.
impose on data analysis design. Programming frameworks can help
improve the correctness and reliability of differentially private programs. However, as we discussed in
\cref{sec:programming-frameworks}, most of the programming
frameworks allow one to assemble basic primitives as building blocks
that are considered trusted. This situation is less than ideal, since
bugs in these primitives can affect the correctness of the overall
data analyses. 
In order to help
alleviate this problem, researchers have developed several
verification and testing tools specifically designed to improve the
correctness and reliability of differentially private primitives.
In this section, we briefly review some of them, focusing on the techniques that they are based on.
All these methods have to cope with the fact that black-box testing for differential privacy, as well as automated verification of differential privacy, are inherently difficult in the worst case~\cite{gilbert2018property,GaboardiNP20,BunGG22}. 

\subsection{Tools for testing differential privacy implementations}
Testing is a powerful tool to improve the reliability and correctness of software. 
Following this intuition, several testing methods have been studied and implemented in order to 
increase confidence in claims of differential privacy for data analysis implementations. 
Early work on testing for differential privacy~\cite{JhaR11,DixitJRT13} reduced this problem to testing Lipchitz properties of data analyses, while subsequent work designed methods specific for differential privacy.
\tool{StatDP}~\cite{DingWWZK18} provides a testing framework based on hypothesis testing and counterexample generation to find violations in data analyses that claim to be differentially private. This testing framework considers programs as black box but also provides the option to use program analysis methods to estimate parameter values that can create violations. 
Following similar ideas, \tool{GoogleSQL}~\cite{WilsonZLDSG20} includes a stochastic tester in order to check whether some basic mechanism is differentially private. This tester is useful to analyze mechanisms of a restricted class, corresponding to aggregation functions, rather than end-to-end applications.  
\tool{DP-Finder}~\cite{BichselGDTV18} uses testing and numerical optimization techniques to prove lower bounds on the privacy parameters of specific programs. This method can be also used to find violations to differential privacy in implementations.  
\tool{CheckDP}~\cite{WangDKZ20} combines statistical testing with symbolic solving in order to guarantee data analyses are differentially private; when the differential privacy guarantee is violated, this approach provides counterexamples. 
\tool{DPCheck}~\cite{ZhangRHP020} uses a combination of randomized testing and symbolic execution to identify violations of differential privacy in an automated way. 
\tool{DP-Sniper}~\cite{BichselSBV21} uses learning techniques to train a classifier to distinguish whether the result of a mechanism comes from a dataset or an adjacent one, and when possible to transform this classifier into a counterexample exhibiting a violation of differential privacy. \tool{DP-Sniper} is also able to identify violations due to floating-point arithmetic.

\subsection{Tools for the verification of differential privacy}
\label{subsec:verification}
Randomized testing as discussed above can help improve the reliability of differential
privacy components.  However, when the input space of these components
is large, it is difficult for randomized testers to offer strong
guarantees in terms of coverage, and the uncovered input space can
still hide bugs in the design and implementations of these
components. 
For this reason, several groups of
researchers have also developed formal verification techniques which
can be used to mathematically prove the design and implementations of
these components to be correct.
An advantage of these approaches is that the verification can happen 
before the program is run, limiting the risks for the actual data.  

A first approach in this line of work is based on the use of \emph{program
logics} and \emph{interactive proof assistants}. These are formal tools that
have been developed in order to formally verify properties of programs and their
implementation. The idea behind these tools is that a user needs to provide a program, a specification that the program needs to satisfy, and a formal proof, in the form of a computer script, that the program meets the specification. The tool then checks that the formal proof is correct and that indeed the program respects the specification. In the case of differential privacy, the specification is that the program is differentially private. Several works have followed this approach~\cite{BartheKOB12,BartheDGKB13,BartheGAHKS14,BartheGGHS16,BartheFGGHS16,AlbarghouthiH18} building on different program logics and proof assistants.

Another approach in this line of work is based on the use of advanced \emph{type systems}, \emph{symbolic execution} and \emph{constraint solvers}. In this approach, a user needs to provide a program and its specification. The specification is usually in the form of a type specification in some rich language of types that also provides the guarantee that a program is differentially private. However, the user does not need to provide a formal proof connecting the program with its specification. Instead, this formal proof is synthesized by a type checker combined with constraint solving. 
In \cref{subsec:privacy-calculus}, we already discussed \tool{Fuzz}~\cite{reed2010distance}, which was one of the first tools in this area. 
\tool{Fuzz} is limited in expressivity: it can only type programs whose sensitivity is a constant. 
Several works have extended this approach~\cite{gaboardi2013linear,BartheGAHRS15,ZhangK17,WangDWKZ19,WangDKZ20,FarinaCG21} and designed more and more expressive tools. 
The high level of confidence and formal guarantees that these tools provide goes beyond what can be usually achieved by code inspection and randomized testing. On the other hand, those tools are often difficult to use and often require expertise in formal verification or programming-language design, as well as differential privacy.

\newpage
\pagenumbering{arabic}
\renewcommand{\thepage} {B--\arabic{page}}

\bibliography{biblio}
\bibliographystyle{alpha}
\end{document}